\documentclass[nonamebreak]{aa}
\usepackage{mathptmx} 
\usepackage[T1]{fontenc}
\usepackage{savesym} 
\savesymbol{tablenum}
\usepackage{natbib}
\usepackage{graphicx}
\usepackage{amsmath}
\usepackage{amssymb}
\usepackage{epstopdf}
\usepackage{booktabs}		
\usepackage{multirow}	
\usepackage[caption=false]{subfig}		
\usepackage[separate-uncertainty=true]{siunitx}	
\restoresymbol{SIX}{tablenum} 
\usepackage{bm}	
\usepackage{lmodern}	
\usepackage{enumitem}
\usepackage{tabulary}
\usepackage{color}
\usepackage{pgffor}
\usepackage{epstopdf}
\usepackage{footmisc}
\usepackage{wasysym}
\usepackage[colorlinks=true,bookmarks=true,backref=true,allcolors=blue]{hyperref}

\bibpunct{(}{)}{;}{a}{}{,} 

\newcommand{\tup}[1]{\textup{#1}}
\newcommand{\msun}{$ \text{M}_\odot $}

\newcommand{\tw}{\textwidth}
\newcommand{\tht}{\textheight}
\newcommand{\minus}{^{-1}}

\newcommand{\dg}{^{\circ}}
\definecolor{purple}{cmyk}{0,1.0,0,0.6}

\defcitealias{bourdin:SZimaging}{B15}
\defcitealias{bourdin:pressure}{B17}

\begin{document}
\title{Spectral imaging of the thermal Sunyaev--Zel'dovich effect \\ in X-COP galaxy clusters: method and validation}
\author{A.~S.~Baldi\inst{\ref{inst:tov},\ref{inst:sap},\thanks{E-mail: %
	\href{mailto:a.silvia.baldi@gmail.it}{\nolinkurl{anna.silvia.baldi@roma2.infn.it}}}},
	H.~Bourdin\inst{\ref{inst:tov}}, P.~Mazzotta\inst{\ref{inst:tov}},
	D. Eckert\inst{\ref{inst:maxplanck}},
	S. Ettori\inst{\ref{inst:inafbo},\ref{inst:infnbo}},
	M. Gaspari\inst{\ref{inst:princeton},\thanks{Lyman Spitzer Jr. Fellow}},
	M. Roncarelli\inst{\ref{inst:inafbo},\ref{inst:unibo}}
	}
\institute{Universit\`{a} di Roma ``Tor Vergata'', Via della Ricerca Scientifica, I-00133 Roma, Italy \label{inst:tov}
\and Sapienza Universit\`{a} di Roma, Piazzale Aldo Moro 5, I-00085 Roma, Italy \label{inst:sap}
\and Max-Planck-Institut für Extraterrestrische Physik, Giessenbachstrasse 1, 85748 Garching, Germany \label{inst:maxplanck}
\and INAF, Osservatorio di Astrofisica e Scienza dello Spazio, via Pietro Gobetti 93/3, 40129 Bologna, Italy 
\label{inst:inafbo}
\and INFN, Sezione di Bologna, viale Berti Pichat 6/2, 40127 Bologna, Italy \label{inst:infnbo}
\and Department of Astrophysical Sciences, Princeton University, 4 Ivy Lane, Princeton, NJ 08544-1001, USA 
\label{inst:princeton}
\and Dipartimento di Fisica e Astronomia, Università di Bologna, via Gobetti 93/2, I-40129 Bologna, Italy \label{inst:unibo}
}
\titlerunning{Spectral imaging of the tSZ effect in X-COP galaxy clusters}
\authorrunning{A. S. Baldi et al.}
\date{Received ------ / Accepted ------}
\abstract{
	The imaging of galaxy clusters through the Sunyaev--Zel'dovich effect is a valuable tool to probe
	the thermal pressure of the intra-cluster gas, especially in the outermost regions where X-ray 
	observations suffer from photon statistics.
	For the first time, we produce maps of the Comptonization parameter by applying a locally parametric algorithm for 
	sparse component separation to the latest frequency maps released by \textsl{Planck}.
	The algorithm takes into account properties of real cluster data through the two-component modelling of the spectral
	energy density of thermal dust, and the masking of bright point sources.
	Its robustness has been improved in the low signal-to-noise regime, thanks to the 
	implementation of a deconvolution of \textsl{Planck} beams in the chi-square minimisation of each
	wavelet coefficient.
	We applied this procedure to twelve low-redshift galaxy clusters detected by \textsl{Planck}
	with the highest signal-to-noise ratio, considered in the XMM Cluster Oustkirts Project (X-COP).
	Our images show the presence of anisotropic features, such as small-scale blobs and filamentary substructures
	that are located in the outskirts of a number of clusters in the sample. The significance of their detection
	is established via a bootstrap-based procedure we propose here for the first time.
	In particular, we present a qualitative comparison with X-ray data for two interesting systems, namely A2029 and
	RXCJ1825. Our results are in agreement with the features detected in the outskirts of the clusters in the
	two bands.
}
\keywords{Techniques: image processing -- Galaxies: clusters: general -- Galaxies: clusters: intracluster medium --
Galaxies: clusters: individual: A2029; RXC J1825.3+3026}
\maketitle
\section{Introduction}
\label{sec:introduction}
Clusters of galaxies, which are the largest gravitationally bound structures in the universe, can be studied at
millimetre wavelengths through observations of the Sunyaev--Zel'dovich (SZ) effect \citep{sz:70,sz:72}.
Indeed, as the photons of the cosmic microwave background (CMB) radiation propagate through the hot and ionised
intra-cluster medium (ICM hereafter) of galaxy clusters, they get diffused via inverse Compton scattering by
the free electrons. The consequence of this interaction is that photon energy is re-distributed; therefore, the
observed black body spectrum of the CMB gets distorted.
The spectral distortion produced by the random thermal motion of the electrons is known as the thermal SZ (tSZ) effect.
A remarkable property is its linear dependence on both electron density and temperature, making it a direct probe
of the thermal pressure of the ICM integrated along the observer line of sight.
For this reason, the tSZ effect is extremely useful when investigating the outskirts of galaxy clusters, which host complex 
physical processes associated to the continuous accretion of matter
\citep[see e.g.][for a recent review]{walker:outskirtsreview}.
In this respect, the XMM Cluster Outskirts Project \citep[X-COP, see][]{xcop:presentation}, proposes combining the
tSZ effect with X-ray data at radii $ R_{500}< r < R_{200} $\footnote{Whenever the subscript $ \Delta = 200,500 $ is used, we
	refer to the mass enclosed by a spherical volume of radius $ R_{\Delta} $, whose density is $ \Delta $ times the 
	critical density of the universe at a given redshift $ z $, $ \rho_c(z) = 3 H(z)^2/(8 \pi G) $, being $ H(z) $ the
	Hubble parameter, and $ G $ the gravitational constant},
which are barely accessible to X-ray spectroscopy because of
significant contamination by unresolved astrophysical or instrumental backgrounds.

Another important feature of the tSZ effect, which makes it unique among other observables at microwave frequencies,
is its peculiar spectral signature. Indeed, the observed change of the CMB brightness manifests as a decrement at frequencies
below the zero-frequency $ \nu_0 = \SI{217}{\GHz} $, and as an increment at higher frequencies.
In the non-relativistic approximation \citep{kompaneets:articolo}, both the zero-frequency and the overall spectral
signature do not depend on cluster physical parameters. Therefore, thanks to this characteristic, it is possible to 
disentangle the tSZ signal from other astrophysical sources using multi-frequency data.

The separation of different components contributing to an astrophysical signal is an issue that has been extensively
addressed in the literature with a variety of techniques, and for diverse purposes.
Among the most used methods, there are refinements of the internal linear combination (ILC),
which was first proposed by \citet{eriksen:ilc}.
These methods assume there is no correlation between the different physical components, and they are non-parametric,
which means that they do not rely on any particular model. The contribution from each astrophysical source is estimated 
through a simple linear  combination of the total signal at all the available frequencies. The weighting coefficients for
each frequency band are chosen in such a way that they provide unit response with respect to the component of interest,
while minimising the variance of the reconstructed map.
The increasing amount of microwave data at high angular
resolution, which can be of a few arcmin or arcsec with currently operating instruments
\citep[see e.g.][for a recent review]{mroczkowski:szreview},
and the unprecedented frequency coverage provided by
the \textsl{Planck} satellite \citep{planck:presentation}, motivated the development of several modified ILC algorithms
tailored to the extraction of the tSZ effect.
Relevant examples in this respect are provided by: \citet{remazeilles:cilc}, \citet{hurier:milca} (who propose the
modified internal linear combination algorithm: MILCA), and \citet{hurier:milcann}.
Despite their success, ILC methods for the extraction of the tSZ effect suffer from some drawbacks
\citep[see e.g.][for a discussion]{bobin:compsep}.
For instance, results can be significantly biased if the different foreground sources are correlated, as well as if
they exhibit a spatially variable spectral energy density.
Moreover, ILC methods perform better in the case of Gaussian foreground fluctuations, which is not a realistic hypothesis
for the tSZ effect and the emission from thermal dust.

In addition, the combination of heterogeneous data sets from ground-based and space-based experiments has been explored
through the proposal of more sophisticated techniques, with the aim of exploiting the advantages coming from
observations at different angular resolutions and sensitivities.
For instance, \citet{remazeilles:nilc} propose a method which implements needlet decomposition, while in the recent work
by \citet{aghanim:pact} the authors apply suitable matched multi-filters
\citep[see e.g.][]{herranz:matchingfilterSZ,melin:SPTSZmapping,tarrio:mmf16,tarrio:mmf18}
to data from \textsl{Planck} and from the Atacama Cosmology Telescope \citep{act:presentation}.

Besides non-parametric algorithms, methods based on a prior modelling of the components and on likelihood maximisation
have also been developed \citep[see e.g.][]{feroz:bayesclusters,khatri:parametricSZ}.
In particular, parametric techniques are robust, especially for the mapping of Galactic thermal dust
\citep[see for instance][]{planck:dust2013,meisner:twodust}.
On the other hand, algorithms based on sparse representations allow one to exploit some properties of the foreground
sources, such as their morphological independence or positive normalisation.

In \citet{bourdin:SZimaging} (\citetalias{bourdin:SZimaging} hereafter) the authors propose to link sparse imaging
to a parametric estimate of foregrounds, via the introduction of a spatially-weighted likelihood function through
wavelets and curvelets.
Wavelets are basis functions characterised by a scale (or equivalently by its inverse, that is the resolution) and by a 
position parameter.
Wavelet transforms provide sparse representations of signals characterised by a smooth distribution, apart from localised
discontinuities. Since the ICM seems to hold such properties, wavelet transforms have been widely used in cluster
astronomy, both for X-ray \citep{slezak:waveletxray,vikhlinin:waveletxray,starck:waveletxray} and tSZ imaging 
\citep{pierpaoli:waveletsz,pires:waveletsz}.
Another relevant application for millimetre astronomy is the foreground cleaning of CMB maps proposed by 
\citet{bobin:compsep,bobin:compsep16}, which relies on sparse priors to spatially separate the components.
Curvelet transforms are a higher-dimensional generalisation of wavelet transforms. Indeed, curvelet bases 
are parametrised by two additional quantities, namely direction and elongation \citep{candes:curvelets}.
This property makes them particularly suitable to get sparse representations of images featuring
edges and anisotropies \citep{mallat:waveletbook}, as in the case of shocks and filamentary structures in
the outskirts of galaxy clusters.

For the reasons above, in \citetalias{bourdin:SZimaging} the authors employ wavelet decomposition in
the extraction of the CMB signal and curvelet decomposition to separate the anisotropic signal from the thermal dust
and the tSZ effect.
The algorithm is tested using mock \textsl{Planck} frequency maps from the set of high-resolution
hydrodynamical simulations of galaxy clusters presented in \citet{planelles:sims14} and \citet{rasia:sims14}.

We applied for the first time the spectral imaging algorithm by \citetalias{bourdin:SZimaging}
to the frequency maps from the latest data released by the \textsl{Planck} Collaboration in 2018.
Furthermore, we improved the algorithm by implementing an efficient removal of astrophysical contaminations,
such as the emission from thermal dust and bright point sources. Most importantly, we enhanced its stability by means
of a new deconvolution technique applied to the wavelet coefficients, proposed for the first time in this work.
We show the maps of the tSZ effect for a set of twelve massive and nearby clusters of galaxies selected for X-COP.
We focus in particular on two interesting cases of possibly interacting cluster couples, and we compare our images
to ancillary X-ray data. The significance of the signal is established through a new boostrap-based procedure for
the estimate of the tSZ error maps, which we also used to show the overall enhancement of the algorithm, compared to 
its previous implementation.

The paper is structured as follows.
In Sect.~\ref{sec:dataset} we briefly introduce the cluster data set.
Section~\ref{sec:methods} illustrates the details of the algorithm we used for producing maps of the tSZ effect,
focussing on its new aspects.
In Sect.~\ref{sec:results} we present and discuss our results for the X-COP sample. In particular, we make a comparison
with the implementation of the algorithm presented in \citetalias{bourdin:SZimaging}, and we detail the cases of
cluster A2029 and RXCJ1825.
Finally, we summarise our results and outline future perspectives in Sect.~\ref{sec:conclusions}.
Throughout the paper, we assume a $ \Lambda $CDM cosmological model with $ \Omega_{\Lambda} = 0.7 $, $ \Omega_m = 0.3 $
and $ h = 0.7 $.
\section{Data set}
\label{sec:dataset}
We applied our imaging procedure to the sample of galaxy clusters studied in the XMM Cluster Outskirts Project (X-COP).
The goal of the project is to unveil the thermodynamic and dynamical properties of the ICM in the outermost cluster regions,
combining data in the X-ray and millimetre bands \citep[see][]{tchernin:a2142,xcop:presentation,
xcop:a2319,xcop:hydromassprofiles,xcop:nonthermalpressure,xcop:thermoproperties}.

The main physical properties of the X-COP clusters are summarised in Table~\ref{tab:xcop}.
\begin{table*}
	\centering
	\caption{\small Basic data of X-COP galaxy clusters~\citep[taken from][]{xcop:thermoproperties}.
		Galactic coordinates are taken from the NASA/IPAC Extragalactic Database (\url{https://ned.ipac.caltech.edu}).}
	\begin{tabular}{cccccccc}
		\toprule
		\midrule
		Cluster name & %
		\textsl{Planck} S/N & %
		$ z $ & %
		$M_{500}$ ($ \times 10^{14}$\msun) & %
		$R_{500} $ (kpc) & %
		$\theta_{500}$ (arcmin) & %
		$l $ (deg) & %
		$b $ (deg)\\
		\midrule
		\object{A2319}   & 30.8 & 0.0557 & 7.31 & 1346 & 20.8 & 75.70 & 13.51    \\ 
		\object{A3266}   & 27.0 & 0.0589 & 8.80 & 1430 & 21.0 & 272.13 & -40.13  \\ 
		\object{A2142}   & 21.3 & 0.0909 & 8.95 & 1424 & 14.1 & 44.22 & 48.68    \\ 
		\object{A2255}   & 19.4 & 0.0809 & 5.26 & 1196 & 13.1 & 93.97 & 34.95    \\ 
		\object{A2029}   & 19.3 & 0.0766 & 8.65 & 1414 & 16.3 & 6.44 & 50.53     \\ 
		\object{A85}     & 16.9 & 0.0555 & 5.65 & 1235 & 19.2 & 115.23 & -72.03  \\ 
		\object{A3158}   & 17.2 & 0.0597 & 4.26 & 1123 & 16.3 & 265.05 & -48.93  \\ 
		\object{A1795}   & 15.0 & 0.0622 & 4.63 & 1153 & 16.1 & 33.82 & 77.18    \\ 
		\object{A644}    & 13.9 & 0.0704 & 5.66 & 1230 & 15.3 & 229.93 & 15.29   \\ 
		\object{A1644}   & 13.2 & 0.0473 & 3.48 & 1054 & 19.0 & 304.88 & 45.45   \\ 
		RXCJ1825$^{\tiny \ddag}$ & 13.4 & 0.0650 & 4.08 & 1105 & 14.8 & 58.31 & 18.54  \\
		ZW1215$^{\tiny \ddag}$ & 12.8  & 0.0766 & 7.66 & 1358 & 15.7 & 282.50 & 65.19  \\
		\bottomrule
	\end{tabular}
	\newline\newline\newline
	$^{\tiny \ddag}$\tiny{The full identifiers for these two clusters are \object{RXC J1825.3+3026} and
		\object{ZwCl 1215.1+0400}, respectively.}\hfill
	\label{tab:xcop}
\end{table*}
The sample consists of twelve massive objects located at low to intermediate redshift, namely in the range
$ 0.04 < z < 0.10 $, observed in X-ray by \textsl{XMM-Newton} \citep{xmm:fov,xmm:resolution}
and at millimetre wavelengths by \textsl{Planck}.
As discussed in \citet{planck:PSZ1_early}, the six bands observed with the High Frequency Instrument (HFI) centred at
100, 143, 217, 353, 545, and 857 GHz are those which provide the highest signal-to-noise ratio (S/N hereafter)
for the detection of galaxy clusters.
In particular, the clusters in the X-COP sample have been detected with S/N $ > 12 $.
Their characteristic angular size $ \theta_{500} $, that is the angle subtended by the $ R_{500} $ radius of each cluster,
is larger than 10 arcmin, so that they are
well-resolved by \textsl{Planck}. A list of the HFI beams at all frequencies is reported in Table~\ref{tab:hfibeams}.
\begin{table}
	\centering
	\caption{Frequencies and corresponding beams of \textsl{Planck} HFI
		\citep[from][]{planck:hfibeams2015}.}
	\begin{tabular}{cc}
		\toprule
		\midrule
		Frequency (GHz) & Beam FWHM (arcmin)\\
		\midrule
		100 & 9.69\\
		143 & 7.30\\
		217 & 5.02\\
		353 & 4.94\\
		545 & 4.83\\
		857 & 4.64\\
		\bottomrule
	\end{tabular}
	\label{tab:hfibeams}
\end{table}

We extracted $ 256 \times 256 $ pixel maps centred on each target cluster from the raw full-sky
maps provided in the latest \textsl{Planck} data release\footnote{\textsl{Planck} data are publicly
		available at: \url{https://pla.esac.esa.int}} \citep[see][for details on HFI]{planck:hfi2018}.
Specifically, we retrieved the temperature maps, the associated variance maps and the jackknife maps 
\citep[see][]{planck:hfijk} at each frequency band.
In particular, the latter are used in our procedure to estimate the error, as explained
in detail in Sect.~\ref{subsec:error}.
Sky patches are obtained through a gnomonic re-projection using the \texttt{HEALPix} package
\citep{healpix:presentation}.
Each pixel is 1 arcmin in size, so the maps extend over $ \approx 4.3\dg $, corresponding to
sufficiently large radii (i.e. cluster-centric distances $ r > 3R_{500} $) to allow the detection of substructures in
cluster outskirts.

The signal in the raw temperature maps is corrected at all HFI frequencies by the astrophysical offsets
listed in Table~\ref{tab:offsets}, which account for the cosmic infrared background and the tSZ background.
\begin{table}[b]
	\centering
	\caption{Astrophysical offsets subtracted from each raw HFI map (see text for details).}
	\begin{tabular}{cc}
		\toprule
		\midrule
		Frequency (GHz) & Offset\\
		\midrule
		100 & \num{1.47e-5} K$_\tup{CMB}$\\
		143 & \num{2.32e-5} K$_\tup{CMB}$\\
		217 & \num{7.02e-5} K$_\tup{CMB}$\\
		353 & \num{4.12e-4} K$_\tup{CMB}$\\
		545 & \num{3.41e-1} MJy sr$ \minus $\\
		857 & \num{5.84e-1} MJy sr$ \minus $\\
		\bottomrule
	\end{tabular}
	\label{tab:offsets}
\end{table}
The offset values are computed as described in \citet{planck:dust2013}, from the correlation between
HFI data at 857 GHz and data delivered by the Leiden/Argentine/Bonn (LAB) radio survey of Galactic
HI \citep{lab:4hfioffsets}.
The pixels in the sky that are used to compute this correlation are selected according to the values of both the column
density and the velocity of HI clouds. In order to remove the contamination from galaxy clusters, and to account
for calibration at high frequencies, we also added two more selection criteria. One is based on masking pixels where
the the Comptonization parameter estimated from the maps obtained with the MILCA algorithm \citep{hurier:milca}
exceed the threshold value of \num{e-5}.
The second one accounts for a correction of the CMB using SMICA maps \citep{smica:presentation}
at frequencies between 100 GHz and 353 GHz as templates.
\section{Imaging of the tSZ effect}
\label{sec:methods}
To map the tSZ signal, we used a parametric imaging algorithm featuring wavelet and curvelet decomposition.
This procedure was originally proposed in \citetalias{bourdin:SZimaging} and tested only on mock HFI observations
of interacting clusters.
The first application to real HFI data motivated substantial adaptations and improvements of
its original implementation.
In particular, we used the spectral energy density of a double grey body to model the emission from thermal dust,
and we corrected for residual contamination from both dust and bright point sources.
Another remarkable improvement concerns with the stability of the algorithm, which we obtained
through a new deconvolution procedure proposed here for the first time.
In this Section we first provide a synthetic description of the original algorithm, then we focus
on the aforementioned enhancements.
\subsection{The spectral imaging algorithm}
\label{subsec:imaging}
The novelty of our imaging method consists of combining the parametric approach to component separation
and sparse representations.
This is achieved by finding the parameters which minimise a modified chi-square, accounting for the
wavelet transform of the residuals between the data and the model map at each frequency $ \nu $.
Referring to the $ k$-th pixel in the maps, the residuals can be simply written as
\begin{equation}
	\label{eqn:residuals}
	\text{res}(\nu,k;s) = D_\tup{HFI}(\nu,k)-M(\nu,k;s) \ ,
\end{equation}
where $ D_\tup{HFI} $ represent the HFI data.
The source component maps $ s $, which are the parameters we want to estimate, enter the residuals via
the model map, $ M $. The latter can be written explicitly as
\begin{equation}
	\label{eqn:model}
	M(\nu, k; s) = \sum\limits_{i}^{N_s} f_i(\nu) \ s_i(k) + \eta(\nu,k) \ ,
\end{equation}
that is the sum over the temperature anisotropies produced by the $ N_s $ physical sources to separate, $ s_i $,
multiplied by the corresponding spectral energy density $ f_i(\nu) $, plus the instrumental noise, $ \eta $.

The dominating astrophysical sources in the frequency range covered by \textsl{Planck} HFI,
apart from the CMB, are the tSZ effect and the Galactic thermal dust \citep[see for instance][]{planck:foregrounds}.
The spectral energy density of the tSZ effect is treated in the non-relativistic case, so that
\citep[e.g.][]{birkinshaw:szreview}
\begin{equation}
	\label{eqn:ftsz}
	f_\tup{tSZ}(\nu) = \frac{h_p \nu}{k_B T_\tup{CMB}}\times%
	\coth \left(\frac{h_p \nu}{2 k_B T_\tup{CMB}} \right) - 4 \ ,
\end{equation}
where $ h_p $ is the Planck constant, $ k_B $ is the Boltzmann constant, and $ T_\tup{CMB} = (2.725 \pm 0.001)$ K
\citep{mather:tcmb} is the CMB temperature, which is constant at all frequencies.
The amplitude of the tSZ signal is given by the Compton $ y $-parameter \citep[see e.g.][]{rephaeli:articolo}:
\begin{equation}
	\label{eqn:y}
	y = \frac{\sigma_T}{m_e c^2} \int_\tup{los} p(l) \ dl \ ,
\end{equation}
where the integral is calculated along the line of sight (los), being $ \sigma_T $ the Thomson cross section,
$ m_e $ the electron mass, $ c $ the speed of light, and $ p(l) $ the thermal pressure of the ICM.
The spectral modelling of Galactic thermal dust (td subscript) is presented in detail in 
Sect.~\ref{subsec:imaging_novelties},
since it represents a major change in the algorithm.

The residuals in Eqn.~\eqref{eqn:residuals} can be rewritten in terms of their wavelet transform as
\citep[see e.g.][]{mallat:waveletbook}
\begin{equation}
	\label{eqn:residuals15}
	\begin{split}
	\text{res}_{\Psi}(\nu,k;s) &= \sum_n^{N_\tup{pix}} \bar a_{j_0,n}(\nu;s) \ \Phi_{j_0,n}(k) \ +\\
	&+ \sum_{j=j_0}^{N_\tup{scales}} \sum_n^{N_\tup{pix}} a_{j,n}(\nu;s) \ \Psi_{j,n}(k) \ ,
	\end{split}
\end{equation}
where $ j $ and $ n $ give the dilation and translation of the wavelet basis function $ \Psi $, respectively, being
$ 2^j $ the wavelet scale. In our case, the wavelet basis function is a B$_3 $ spline, and $ \Phi $ is the dual scaling
function of $ \Psi $ at the scale $ j_0 $. This scale corresponds to the approximation level -- also called last smooth --
which encodes signal information at the lowest resolution.
The wavelet coefficients in Eqn.~\eqref{eqn:residuals15} are given by
\begin{align}
	&\bar a_{j_0,n}(\nu;s) = \sum_m^{N_\tup{pix}} \text{res}(\nu,m;s) \ \Phi^*_{j_0,n}(m) \ ; \label{eqn:coeffresiduals}\\
	&a_{j,n}(\nu;s) = \sum_m^{N_\tup{pix}} \text{res}(\nu,m;s) \ \Psi^*_{j,n}(m) \ , \label{eqn:coeffresidualsdetail}
\end{align}
for the approximation and detail levels, respectively.
In order to ensure normalisation and positivity when weighting the residuals, the wavelet kernel $ \Psi $ is
split in its positive and negative parts, $ \Psi_{+} $ and $ \Psi_{-} $ (see fig.~1 of \citetalias{bourdin:SZimaging}
for reference).
This yields the minimisation of two separate chi-squares, which can be expressed as the sum across all the
frequencies and pixels ($ N_{\nu} $ and $ N_\tup{pix} $, respectively) of such weighted residuals:
\begin{equation}
	\label{eqn:chi2B15}
	\chi^2_{\Psi_{\pm}} = \sum_{\nu}^{N_{\nu}} \sum_{k}^{N_\tup{pix}} \
	\frac{\text{res}_{\Psi_{\pm}}^2(\nu,k;s)}{\sigma_\tup{HFI}^2(\nu,k)} \ ,
\end{equation}
where $ \sigma^2_\tup{HFI} $ is the HFI variance map.
Therefore, the estimate of the component maps is given by the half-sum of the results from the two separate
minimisations:
\begin{equation}
	\label{eqn:wtcoeffs}
	\hat s = \frac{1}{2} \left[\underset{s}{\text{argmin}}(\chi^2_{\Psi_{+}}) - 
	\underset{s}{\text{argmin}}(\chi^2_{\Psi_{-}})\right] \ .
\end{equation}
To maximise the efficiency in the recovery of signal anisotropies, we chose a curvelet basis to decompose
the tSZ and dust signals \citep[see e.g.][]{candes:curvelets}, and a wavelet basis for the CMB
\citepalias[see][for more details]{bourdin:SZimaging}. The decomposition has been performed over
four scales\footnote{The maximum number of scales that can be used for
wavelet decomposition is linked to the number of pixels on each side of the image as:
$ N_\tup{scales} = \text{floor} [\log(N_\tup{pix})/\log 2]$ - 4}.
The curvelet transform has been computed following the procedure described in
\citet{starck:curvelets}, that is to say trough ridgelet transforms applied to the bidimensional wavelet bands of
the B$_3 $ spline wavelet transform.
Subsequently, we applied a soft thresholding at $ 1 \sigma $ level to the curvelet and wavelet 
coefficients, in order to keep only the relevant features of the signal.
Eventually, the final maps have been obtained by means of a suitable restoration operator,
which re-combines the approximation coefficients and the detail coefficients together.
Pixels plagued by contaminants, due for instance to a local bad modelling of thermal dust, have been automatically
masked. Specifically, we discarded those pixels where chi-square exceeded a given threshold value, depending on the
wavelet scale. At the same time, we imposed a condition of regularity on the error of the wavelet coefficients, in order
to select regions characterised by a high S/N of the tSZ component.
\subsection{Improvements and new features}
\label{subsec:imaging_novelties}
The three main changes we implemented with respect to the version of the imaging
algorithm presented in \citetalias{bourdin:SZimaging} are detailed in the following subsections.
Two of them concern with a more realistic treatment of astrophysical contaminants. The third one consists of a new
deconvolution technique proposed here for the first time.
\subsubsection{Modelling of thermal dust}
Following \citet{meisner:twodust} \citep[see also][]{bourdin:pressure},
we modelled the thermal dust as a double grey body, by assuming two populations of dust grains, instead of
the idealised case of a single grey body spectrum. Indeed, the latter provides an accurate representation of the
thermal emission from Galactic dust only at frequencies higher than 353 GHz.
The spectral energy density we set for this component is therefore
\begin{equation}
	\label{eqn:meisnergb}
	\begin{split}
	f_\tup{td}(\nu) &=
	f_1 \frac{q_1}{q_2} \left(\frac{\nu}{\nu_0} \right)^{\beta_1} B(\nu;T_1) \ + \\
	&+ (1-f_1) \left(\frac{\nu}{\nu_0} \right)^{\beta_2} B(\nu;T_2) \ ,
	\end{split}
\end{equation}
where the dimensionless constant factors $ f_1 $ and $ q_1/q_2 $ refer to the relative contribution
from the coldest component at temperature $ T_1 $ and the hottest component at temperature $ T_2 $.
The $ \beta_1 $ and $ \beta_2 $ parameters give the slopes of the two different power laws, while
$ B(\nu; T_1) $ and $ B(\nu; T_2) $ are the corresponding Planck functions describing the black body spectra.
In order to get the best-fit parameters of the model described by Eqn.~\eqref{eqn:meisnergb},
we calculated an independent fit to the dust component through a Monte Carlo Markov Chain sampling.
To treat only the signal from the dust, we limited this fit to the pixels in the frequency maps that are located
sufficiently far from the cluster, at radial distances from the centre larger than $ 5 R_{500} $.
The only spatially-variable parameter is the temperature $ T_2 $, which is fixed a priori to the value
determined by a joint fit to \textsl{IRAS} and \textsl{Planck} data, as detailed in \citet{meisner:twodust}.
From this fit we obtain maps of the dust component at all frequencies, which we plug in the model
maps of Eqn.~\eqref{eqn:model}.
\subsubsection{Removal of contamination from thermal dust and point sources}
\label{subsubsec:adaptation}
Despite the two grey body model provides an accurate description of the thermal dust component,
our maps of the Compton $ y $-parameter turned out to be affected by contamination either from diffuse signal
from thermal dust on large spatial scales\footnote{We notice, however, a significant reduction of this
contamination when comparing maps from \textsl{Planck} 2015 data release to those from the 2018 data release},
or from compact point sources, mostly at the frequency of 857 GHz.
To solve these issues, we proceeded as described in the following.

We reconstructed the tSZ images without accounting for the contribution from the 857 GHz frequency
in the last approximation of its discrete wavelet transform, $ \bar a_{j_0,n} $ (see also Eqn.~\eqref{eqn:residuals15}).
On the other hand, we took advantage of all the HFI frequency maps to compute the detail coefficients of
the wavelet transform, $ a_{j,n} $, supposedly insensitive to image gradients below the last approximation scale,
$ j_0 $. This allowed us to  suppress unphysical large-scale gradients in the approximation coefficients of 
some cluster images. These gradients could be possibly related to a residual contamination of the thermal dust signal
by the large-scale variance of other astrophysical sources or of the instrumental noise.

To remove the contamination from point sources, we used the masks in the second \textsl{Planck} catalogue
of compact sources (PCCS2) \citep[see][for details]{planck:compactsources,planck:compactsources2}.
These objects are dusty galaxies and synchrotron emitting in the HFI range at high and low frequencies,
respectively.
Specifically, we combined the gnomonic projection of all the masks at the six HFI frequencies into a unique
mask, after smoothing them to the common resolution of 18 arcmin. To remove the contribution from such sources,
we multiplied this mask by the wavelet kernel used in the chi-square minimisation. We applied this procedure only to
the clusters in our sample which are significantly affected by bright point sources at far infrared frequencies,
namely A3266, A85, and ZW1215.

\subsubsection{Deconvolution procedure}
\label{subsubsec:deconvolution}
As can be seen from Eqn.~\eqref{eqn:chi2B15}, the weighted chi-square mixes the different angular resolutions
of the data. Therefore, a deconvolution must be implemented to restore the signal.
The original version of the algorithm features a Van Cittert deconvolution \citep{vancittert:deconvolution},
regularised via successive projections in the significant curvelet domain.
In other words, the resolution of the model map is adjusted to the resolution of
the data map through iterative corrections of the curvelet coefficients.
However, several tests revealed that this technique hampered the reliability of the signal reconstructed in low
signal regimes, because of the amplification of pixel-sized diverging artefacts across the iterations.

\begin{figure}
	\centering
	\subfloat[Parameter $ s $]{\includegraphics[width=0.23\tw]{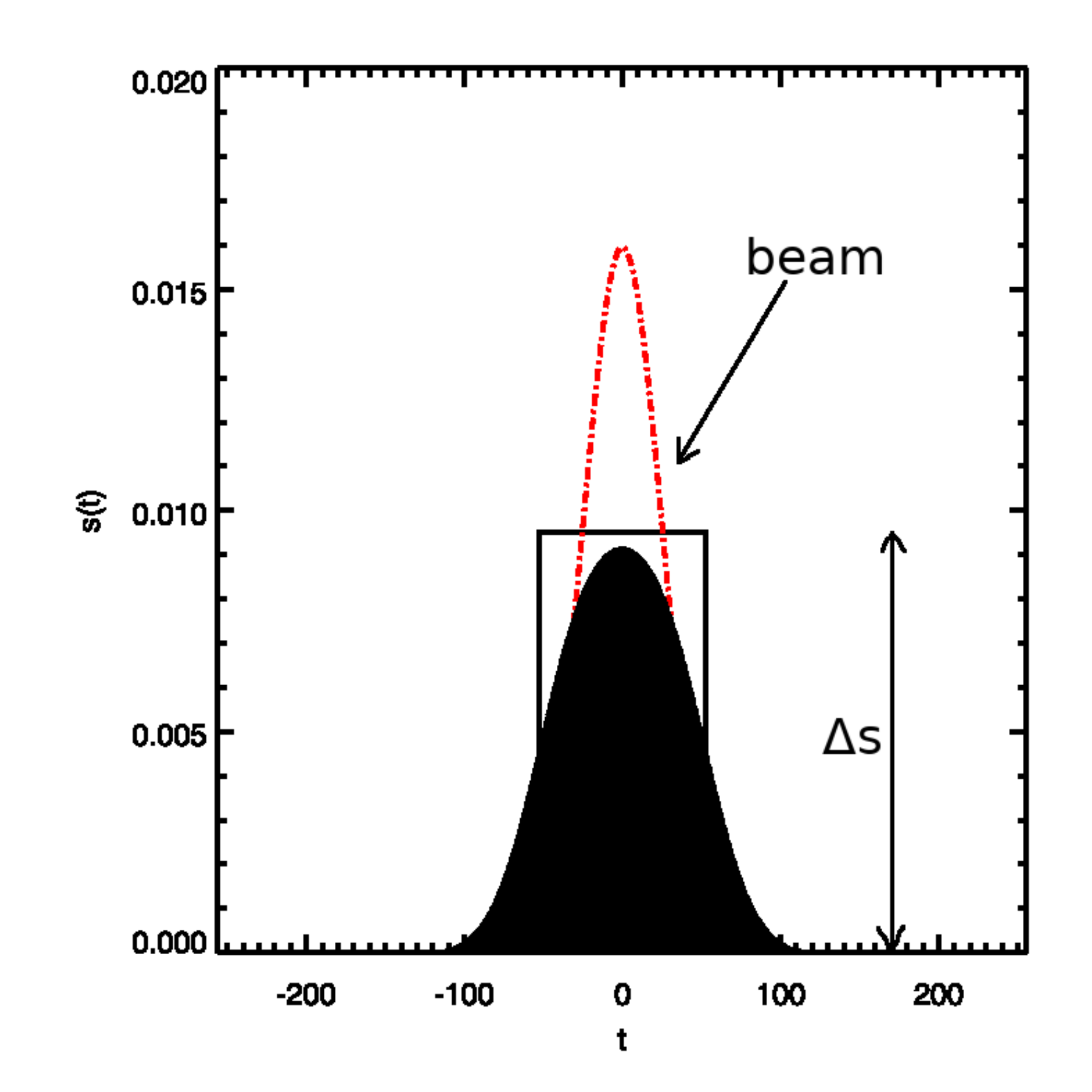}
	\label{fig:parameterdeconv}}
	\subfloat[Wavelet kernel]{\includegraphics[width=0.23\tw]{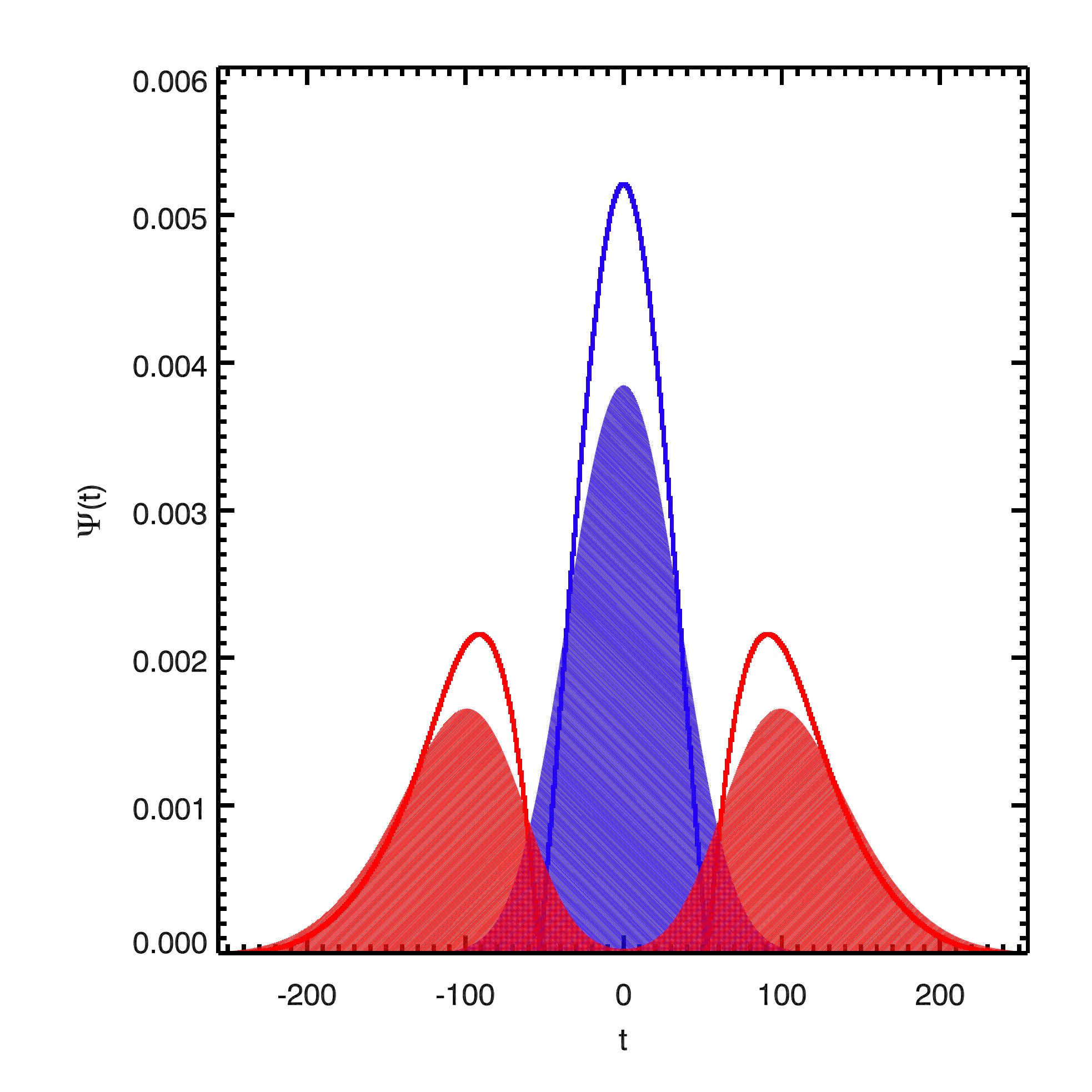}
	\label{fig:kerneldeconv}}
	\caption{Effect of beam-weighting of wavelet coefficients.
	\textbf{Fig.~\ref{fig:parameterdeconv}.} Step-like spatial variation of $ s $ (solid black line).
	The filled black area represents its convolution with the instrumental beam (dashed red line).
	\textbf{Fig.~\ref{fig:kerneldeconv}.} Effective spatial variation of $ s $.
	The blue and red lines represent the absolute value of the positive and negative parts of the wavelet kernel, 
	respectively. The envelopes of the shaded areas coincide with the convolution of the instrumental beam with
	the support of the positive and negative parts of the kernel.}
	\label{fig:newdeconv}
\end{figure}

To avoid this problem, we implemented a new ``wavelet coefficient-wise'' deconvolution of the \textsl{Planck} beams.
Specifically, we deconvolved the amplitude of the single wavelet coefficients, instead of deconvolving the final image.
To this end, we took the full expression of the residuals as in Eqn.~\eqref{eqn:residuals}, and we calculated their
wavelet coefficients as follows (see also Eqn.~\eqref{eqn:coeffresidualsdetail} for a comparison):
\begin{equation}
	\label{eqn:coeffresidualsdeconv}
	\begin{split}
		a_{j,n}(\nu;\Delta s) &= \sum_m^{N_\tup{pix}} \text{res}(\nu,m; \Delta s) \ \Psi^*_{j,n}(m) \\
		&=\sum_m^{N_\tup{pix}} \lbrace D_\tup{HFI} (\nu,m) - \mathcal{B}(\nu) \otimes [H M(\nu,m;\tilde s + \Delta s)+\\
		&+ (1-H) M(\nu,m;\tilde s - \Delta s)]\rbrace \ \Psi^*_{j,n}(m) \ ,
	\end{split}
\end{equation}
(the same holds for the approximation level).
We rewrite the parameter as $ s = \tilde s \pm \Delta s $ in Eqn.~\eqref{eqn:coeffresidualsdeconv}, to
highlight its spatial variation, $ \Delta s $.
Instead of splitting the wavelet kernel in two components, we used the Heaviside step function, $ H $,
that coincides with the positive part of the kernel, while $ (1-H) $ coincides with its negative part.
This operation is equivalent to taking the absolute value of the wavelet function.
To match the angular resolutions of the data and the model, we convolved the model maps at each frequency with the
\textsl{Planck} beam, $ \mathcal{B}(\nu) $, which can be approximated with a Gaussian having the FWHM listed in 
Table~\ref{tab:hfibeams}.
The effect of such a convolution is that of correlating the contributions from the
negative and the positive components of the kernel (see also Fig.~\ref{fig:newdeconv}).
Thus, the final estimate of the parameters takes into account the results from the single decompositions.
Besides allowing us to deconvolve spatial variations of the model maps, the weighting by the absolute value of the wavelet
kernel within a single chi-square minimisation increases the support of the kernel itself, as can be seen in 
Fig.~\ref{fig:kerneldeconv}. This is equivalent to enlarging the size of the sky region where we compute the chi-square, 
thus enhancing the S/N.
\subsection{Procedure for error assessment}
\label{subsec:error}
To associate a statistical error to the estimate of the component maps, we
used a bootstrap procedure that allowed us to simulate $ N_\tup{tot} $ sets of HFI maps.
More specifically, we followed the steps detailed below.

Firstly, we generated $ N_\tup{tot} = 100 $ Monte Carlo realisations of the noise at each HFI frequency,
$ \eta_u(\nu) $, with $ u = 1,\dots, N_\tup{tot} $ referring to the $ u $-th extraction.
We chose the value of $ N_\tup{tot} $ as a trade-off between the computational time needed
to perform wavelet and curvelet transforms and the statistical significance.
We assumed the instrumental noise to be Gaussian and spatially correlated, and we
imposed the noise maps to have the same power spectrum as the jackknife maps.
Subsequently, the HFI raw data at each frequency, $ D_\tup{HFI}(\nu) $, have been denoised through a simple
wavelet-based procedure, by calculating the wavelet transform on three scales, and soft-thresholding the coefficients
at $ 1.5 \sigma $. This procedure relies on spatially-variant thresholds for each wavelet band, which have been 
preliminarily inferred by computing the variance of the coefficients across wavelet transforms of the noise maps,
$ \eta_u(\nu) $.
Lastly, the mock data have been obtained as the summation of the denoised maps,
$ D_\tup{HFI,den}(\nu) $, and the $ u $-th noise realisation, that is
\begin{equation}
\label{eqn:mockhfi}
{D_\tup{HFI}}_u(\nu) = D_\tup{HFI,den}(\nu) + \eta_u(\nu) \ .
\end{equation}
We used such $ N_\tup{tot} $ synthetic data sets as input to the imaging algorithm to obtain the following
vectors of maps:
$ \bm s_\tup{tSZ} = (s_\tup{tSZ}^1, \dots , s_\tup{tSZ}^{N_\tup{tot}}) $,
$ \bm s_\tup{td} = (s_\tup{td}^1, \dots , s_\tup{td}^{N_\tup{tot}}) $,
$ \bm s_\tup{CMB} = (s_\tup{CMB}^1, \dots , s_\tup{CMB}^{N_\tup{tot}}) $.
The standard deviations of all the bootstrap sets, namely $ \text{std}(\bm s_\tup{tSZ})$,
$ \text{std}(\bm s_\tup{td})$, and $ \text{std}(\bm s_\tup{CMB})$, represent our error estimates
for each component.
In the following, we label as $ \sigma_y = \text{std}(\bm s_\tup{tSZ}) $ the error for the tSZ signal only.
The level of significance of the anisotropies (such as blobs and filaments) we find in the tSZ images of each 
cluster can be assessed by computing the ratio $ y/\sigma_y $, for a given minimum value of $ \sigma_y $,
which represents an `effective' S/N.
\section{Results and discussion}
\label{sec:results}
\begin{figure*}[th]
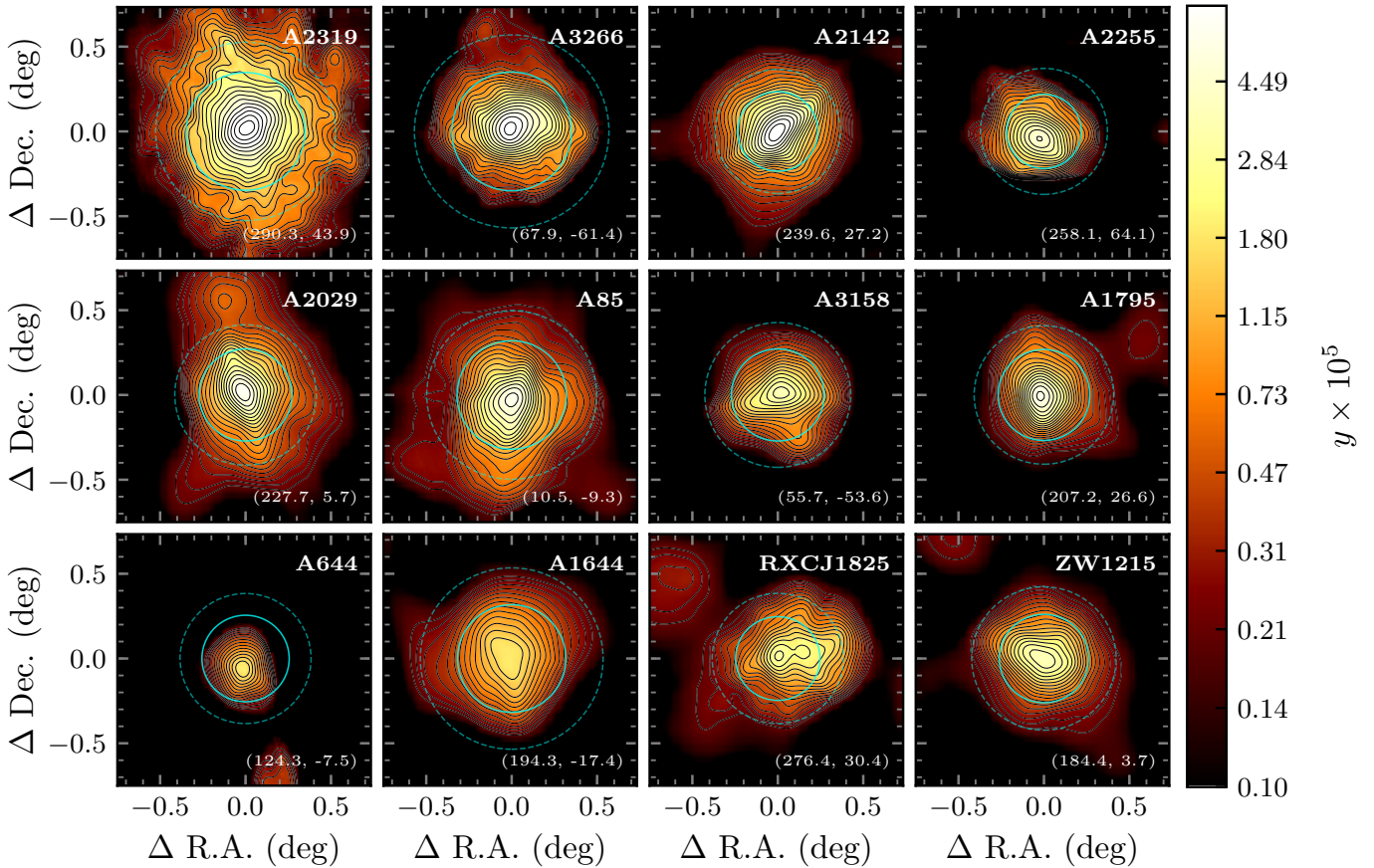

	\centering
	\includegraphics[width=\tw]{{{unique_maps_4scales_128_log_wcsPLANCK2018}}}
	\caption{Compton $ y $-parameter of X-COP clusters from our spectral imaging.
		The solid and dashed cyan circles on top of the images mark the $ R_{500} $ and $ R_{200} $ radii, respectively.
		The colour scale is logarithmic and contours are logarithmically spaced by a factor of $ \sqrt[4]{2} $,
		starting from \num{2e-6}.
		The J2000 Equatorial coordinates of the cluster centres are reported in the bottom right corner of each map.}
	\label{fig:ymaps}
\end{figure*}

We show in Fig.~\ref{fig:ymaps} the zoomed maps of the Compton $ y $-parameter we obtained from the application of the
procedure described in Sect.~\ref{sec:methods} to the HFI maps of the X-COP clusters.
In the majority of cases, the algorithm highlights the presence of blobs or extended filamentary structures at
radial distances $ r \gtrsim R_{500} $ from the cluster centre.
The average minimum values of the Compton $ y $-parameter we can recover from the faintest resolved
structures are \num{2.0e-6} and \num{4.2e-6}, with a minimum significance of $ 3 \sigma_y $ and $ 5 \sigma_y $, respectively.
This represents an improvement of a factor of 2.5 in sensitivity with respect to the version of the algorithm
presented in \citetalias{bourdin:SZimaging}, where the minimum signal for a $ 3 \sigma $ detection is $ y = \num{5e-6} $.
It is important to stress that, thanks to the deconvolution, the algorithm is capable to recovering the elongated
structure of the signal in the cluster central regions (see e.g. the cases of A2142 and A1644).

A particular cluster case which is worth mentioning is that of A2319, which is known to be a complex merging system
that behaves as an outlier with respect to the other clusters in the sample \citep[see][]{xcop:a2319,
xcop:nonthermalpressure,xcop:thermoproperties}.
The tSZ map of this object shows a number of blobs located in the virial region, which may be due to either substructures
or clumpy ICM patches. The detailed study of these features and of their impact on the thermodynamic properties of this
system will be addressed in a forthcoming dedicated paper.

In the following, we first perform a comparison with the results from the version of the algorithm presented in
\citetalias{bourdin:SZimaging} for cluster A2319. Subsequently, we briefly discuss a couple of two possibly
interacting systems, namely A2029 and RXCJ1825.
\subsection{Comparison with the previous version of the algorithm}
We illustrate here the improvements introduced by the adaptation of our procedure to real cluster data, and by the
wavelet coefficient-wise deconvolution. In particular, in order to highlight the impact of removing the highest
frequency from the approximation coefficients (see Sect.~\ref{subsubsec:adaptation}), we adopted a double grey
body model for thermal dust also in the procedure by \citetalias{bourdin:SZimaging}.
To perform a consistent comparison, we applied the two versions of the algorithm to the same HFI maps.
The wavelet decomposition has been performed here over three scales, and coefficients have been thresholded at $ 1\sigma $.

Fig.~\ref{fig:a2319oldnew} shows the maps of the Compton $ y $-parameter from
the old and new versions of the algorithm (see Figs.~\ref{fig:a2319old} and~\ref{fig:a2319new}, respectively),
in the case of cluster \object{A2319}.
\begin{figure}[h]
	\centering
	\subfloat[Van Cittert deconvolution]{\includegraphics[width=0.33\tw]{{{a2319_vc_sz}}}
		\label{fig:a2319old}}\\
	\subfloat[New deconvolution]{
		\hspace{-0.10cm}\includegraphics[width=0.335\tw]{{{a2319_sz}}}
		\label{fig:a2319new}}\\
	\subfloat[HFI at 857 GHz]{
		\hspace{0.02cm}\includegraphics[width=0.33\tw]{{{a2319_800}}}
		\label{fig:a2319857GHz}}
	\caption{Region centred on cluster A2319.
		\textbf{Fig.~\ref{fig:a2319old}.} Compton $ y $-parameter reconstructed with Van Cittert deconvolution.
		\textbf{Fig.~\ref{fig:a2319new}.} Compton $ y $-parameter reconstructed with the
		wavelet coefficient-wise deconvolution.
		The colour scale is logarithmic and contours are logarithmically spaced by a factor of $ \sqrt[4]{2} $.
		\textbf{Fig.~\ref{fig:a2319857GHz}.} Raw HFI map at 857 GHz.}
	\label{fig:a2319oldnew}
\end{figure}
This object is particularly suitable to show the effect of removing the 857 GHz channel from the last smooth
coefficients.
Indeed, in Fig.~\ref{fig:a2319old} it is possible to see a large-scale gradient to the left-hand side of the image
at $ y \approx \num{6e-6} $. Such a contaminant is likely due to a residual
signal from thermal dust, as can be seen from the comparison with the HFI map at 857 GHz, shown in
Fig.~\ref{fig:a2319857GHz}. This residual signal is significantly removed with the new adaptations,
as demonstrated in the map in Fig.~\ref{fig:a2319new}.

Most notably, Fig.~\ref{fig:a2319oldnew} highlights how the two deconvolution techniques have a different impact on
the final maps.
Indeed, the result in Fig.~\ref{fig:a2319old} is produced with the Van Cittert deconvolution using three
iterations, and with the convergence parameter $ \alpha $ set to 0.25 (see eq.~(9) of \citetalias{bourdin:SZimaging}).
It can be seen that this map shows a number of pixel-sized outliers, which plague regions where
the signal is $ y \lesssim \num{7.5e-6} $.
Such artefacts are due to a diverging amplification of the curvelet coefficients, as a consequence of the
iterative nature of this deconvolution algorithm.
The map in Fig.~\ref{fig:a2319new} shows instead that the result from the new wavelet coefficient-wise
deconvolution is cleaner, and the mildly ellipsoidal shape of the
signal in the centre is still well recovered.
It is worth noting that both techniques are sensitive to the basis functions used to project the deconvolved signal.
To be specific, Van Cittert method is a regularisation applied to curvelet coefficients, therefore it is particularly
suitable to highlight ellipsoidal and elongated features. The wavelet coefficient-wise deconvolution, on the contrary,
is applied to wavelet coefficients, thus it provides a better localisation and detection of peaks.
This can be seen in in Fig.~\ref{fig:a2319new}, where the signal in the cluster centre is recovered more efficiently, with
respect to Fig.~\ref{fig:a2319old}.
Another important element that improves peak localisation is the larger kernel
support which enters the chi-square minimisation in the new deconvolution (see Sect.~\ref{subsubsec:deconvolution}
and Fig.~\ref{fig:newdeconv}). Indeed, this feature allows the reconstruction of
the signal with a higher S/N, since in this case the wavelet coefficients are computed and deconvolved on a
larger sky region.

\begin{figure}[t]
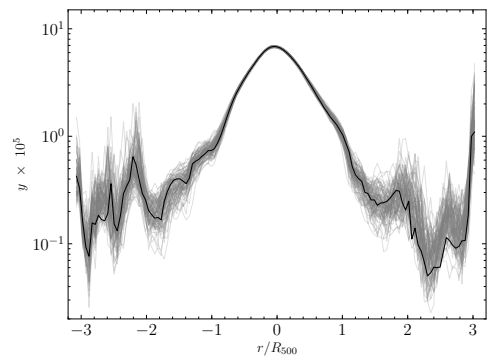
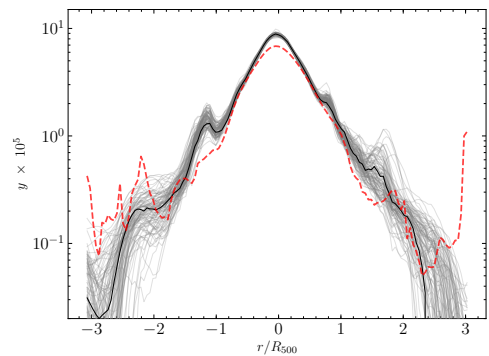

	\centering
	\subfloat[Van Cittert deconvolution] %
	{\includegraphics[width=0.35\tw]{{{a2319_bootstrap_old_reproj_vertical_avg10_log}}}
		\label{fig:a2319boot_old_vert}}\qquad
	\subfloat[New deconvolution] %
	{\includegraphics[width=0.35\tw]{{{a2319_bootstrap_new_reproj_vertical_avg10_log}}}
		\label{fig:a2319boot_new_vert}}
	\caption{Compton $ y $-parameter profiles of cluster A2319.
		The solid black curve of Fig.~\ref{fig:a2319boot_old_vert} is superimposed as a dashed red line to the cuts in
		Fig.~\ref{fig:a2319boot_new_vert} for comparison purposes (see text for details).}
	\label{fig:boot_oldnew}
\end{figure}

To further illustrate the differences between the two deconvolutions, we show as solid black lines in 
Fig.~\ref{fig:boot_oldnew} the profiles of the Compton $ y $-parameter. Each profile has been extracted as the average
vertical cut within a 10 arcmin-wide band, passing through the centre of the corresponding tSZ image in 
Fig.~\ref{fig:a2319oldnew}.
The superimposed light grey lines represent the same profiles from each of the $ N_\tup{tot} = 100 $ different maps of
the Comptonization parameter obtained with the bootstrap procedure (see Sect.~\ref{subsec:error}).
It can be seen from Fig.~\ref{fig:a2319boot_new_vert} that the non-iterative nature of the new deconvolution improves
significantly the stability in low-signal regimes, allowing one to reach a minimum level of $ y $ of the order of \num{e-6}.
On the other hand, the diverging artefacts which hampered the reliability of the detection of substructures in the
outskirts with the Van Cittert deconvolution, are clearly visible in several realisations of the signal
in Fig.~\ref{fig:a2319boot_old_vert}.
Moreover, the dispersion of the bootstrap profiles at radii $ r \gtrsim 2 R_{500} $ is on average 50 per cent lower
with the new procedure, resulting into a lower error in the reconstructed signal in these regions.
The different characterisation of the central peak is also clear from Fig.~\ref{fig:a2319boot_new_vert}, where the
value recovered with the Van Cittert deconvolution is $ \approx 23 $ per cent lower with respect to the value obtained
with the new one based on wavelet coefficients. Such a difference is due to the joint effect of the larger
S/N provided by the absolute value of the wavelet kernel, and of the limited number of iterations (set to 3) used
in the Van Cittert deconvolution. The latter has been arbitrarily chosen as a trade-off between a reliable recovery of the
peak, and the avoidance of divergences. Therefore, it has a non-negligible impact on the final results; on the contrary,
the new deconvolution does not rely on any arbitrary parameter.
From the comparisons reported above, we can conclude that the new version of the algorithm does provide a more reliable
and stable reconstruction of the Comptonization parameter.
\subsection{The cases of A2029 and RXCJ1825}
Clusters A2029 and RXCJ1825 are two interesting targets for testing the algorithm.
Among the X-COP clusters, they have been detected by \textsl{Planck} with intermediate (19.3) and low (13.4) S/N,
respectively.
A2029 has been widely studied in X-rays \citep[see e.g.][]{lewis:a2029,clarke:a2029radiox,bourdin:Ximaging,walker:a2029};
on the contrary, RXCJ1825 has been poorly investigated in this band since its discovery \citep{ebeling:rxc1825discovery}.
Both these clusters may be interacting with two known neighbouring systems, as suggested in
\citet{planck:interactingclusters}. Nevertheless, given the low significance of the data, no further analysis on
the tSZ signal from possible connecting filaments has been explored in their work.

\subsubsection{A2029}
\begin{figure}
	\centering
	\subfloat[A2029 from \textsl{ROSAT}/PSPC] %
	{\includegraphics[height=0.23\tht]{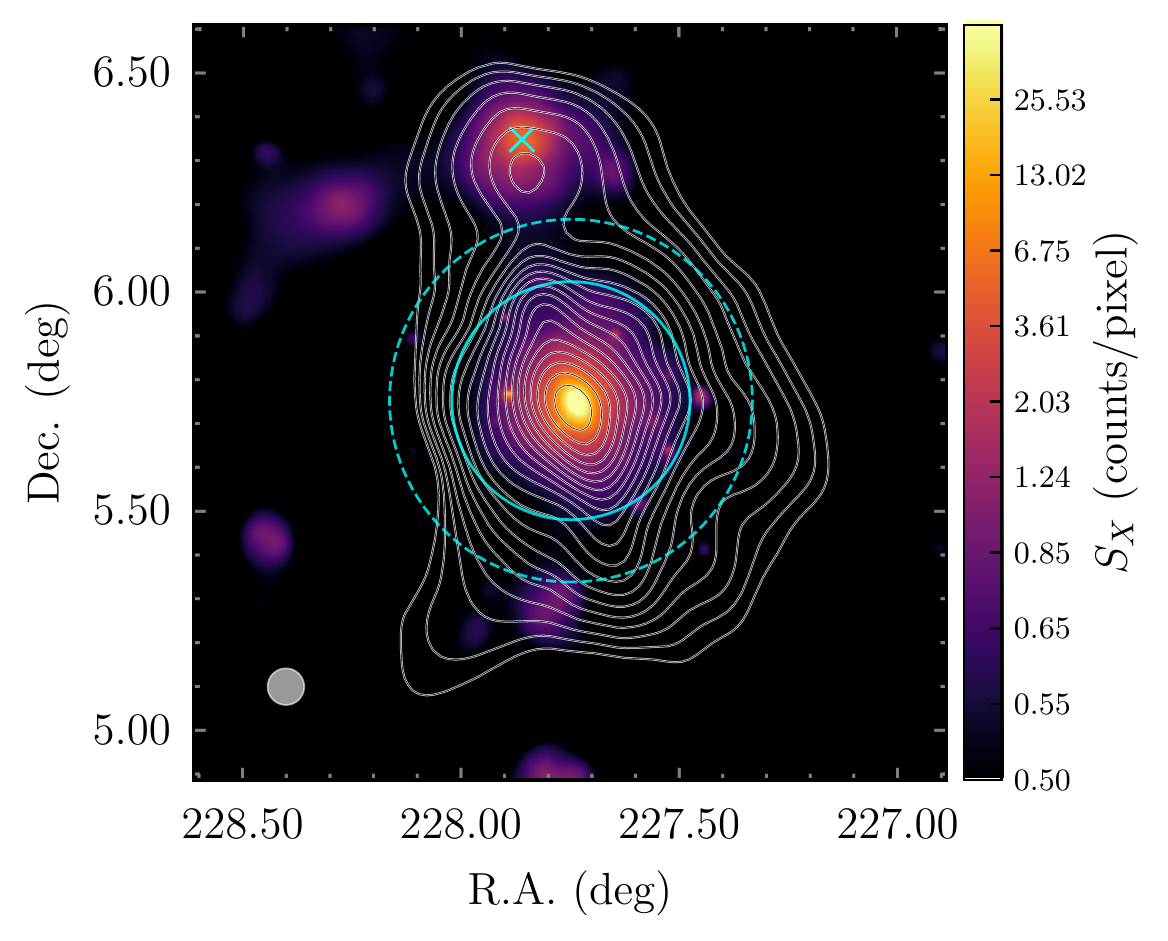} %
		\label{fig:a2029_XSZ}} \quad
	\subfloat[RXCJ1825 from \textsl{XMM-Newton}] %
	{\includegraphics[height=0.23\tht]{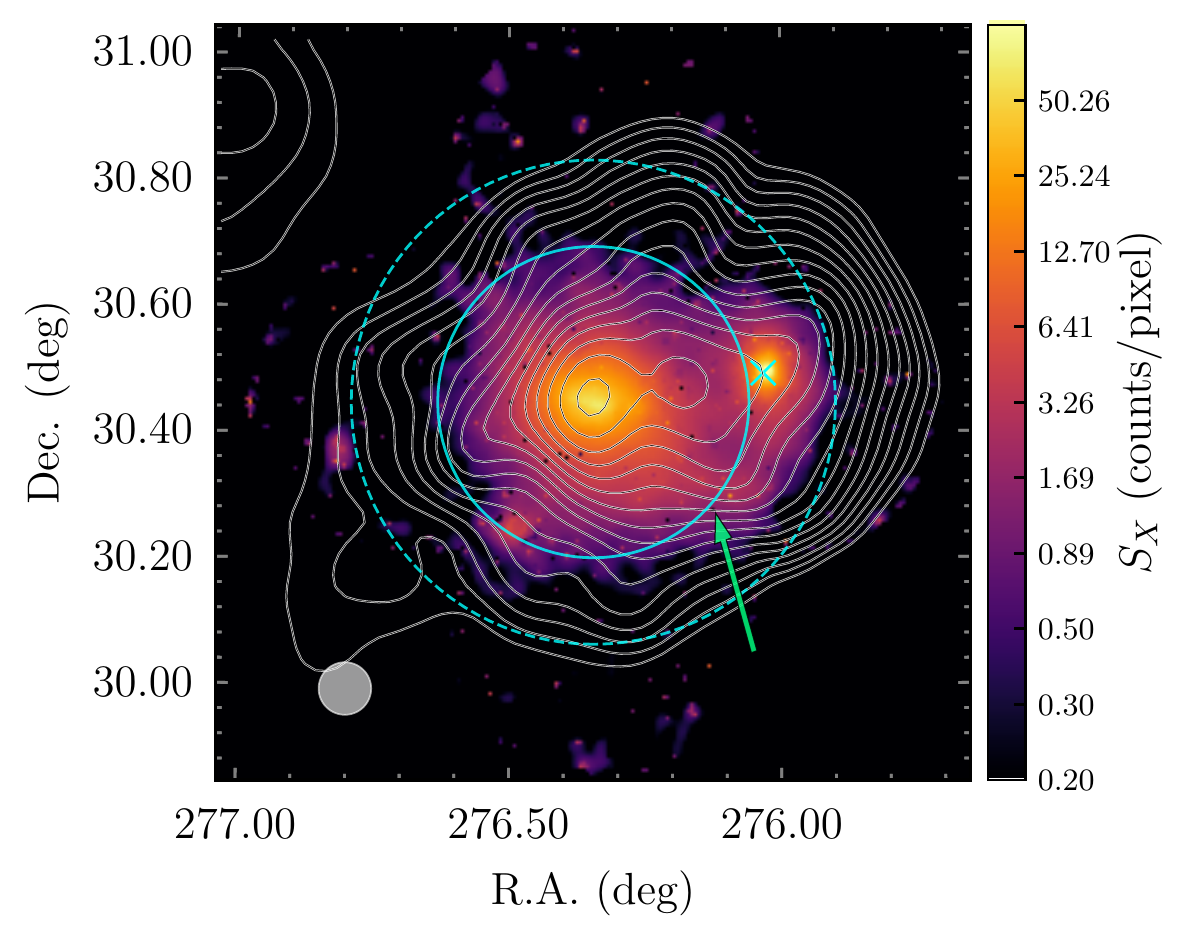}
		\label{fig:rxc1825_XSZ}}
	\caption{X-ray and tSZ signals of clusters A2029 and RXCJ2815.
		The maps show the vignetting-corrected and background-subtracted surface brightness in the X-ray energy band
		0.5-2.5 keV. The superimposed contours represent the Compton $ y $-parameter from our algorithm.
		The solid and dashed cyan circles on top of the images are drawn at $ R_{500} $ and $ R_{200} $, respectively.
		The shaded circles in the bottom left corner of the maps represent a 5 arcmin beam.
		\textbf{Fig.~\ref{fig:a2029_XSZ}.} Cluster A2029 mapped by \textsl{ROSAT}/PSPC.
		The cyan cross indicates the position of the neighbour cluster A2033.
		\textbf{Fig.~\ref{fig:rxc1825_XSZ}.} Cluster RXCJ1825 mapped by \textsl{XMM-Newton}.
		The cyan cross marks the position of cluster CIZAJ1824, while the green arrow identifies a significant
		elongation which may be associated to stripped gas from a nearby group of galaxies.}
	\label{fig:XSZ}
\end{figure}

\begin{figure}[t]
	\centering
	\subfloat[A2029]{\includegraphics[height=0.23\tht]{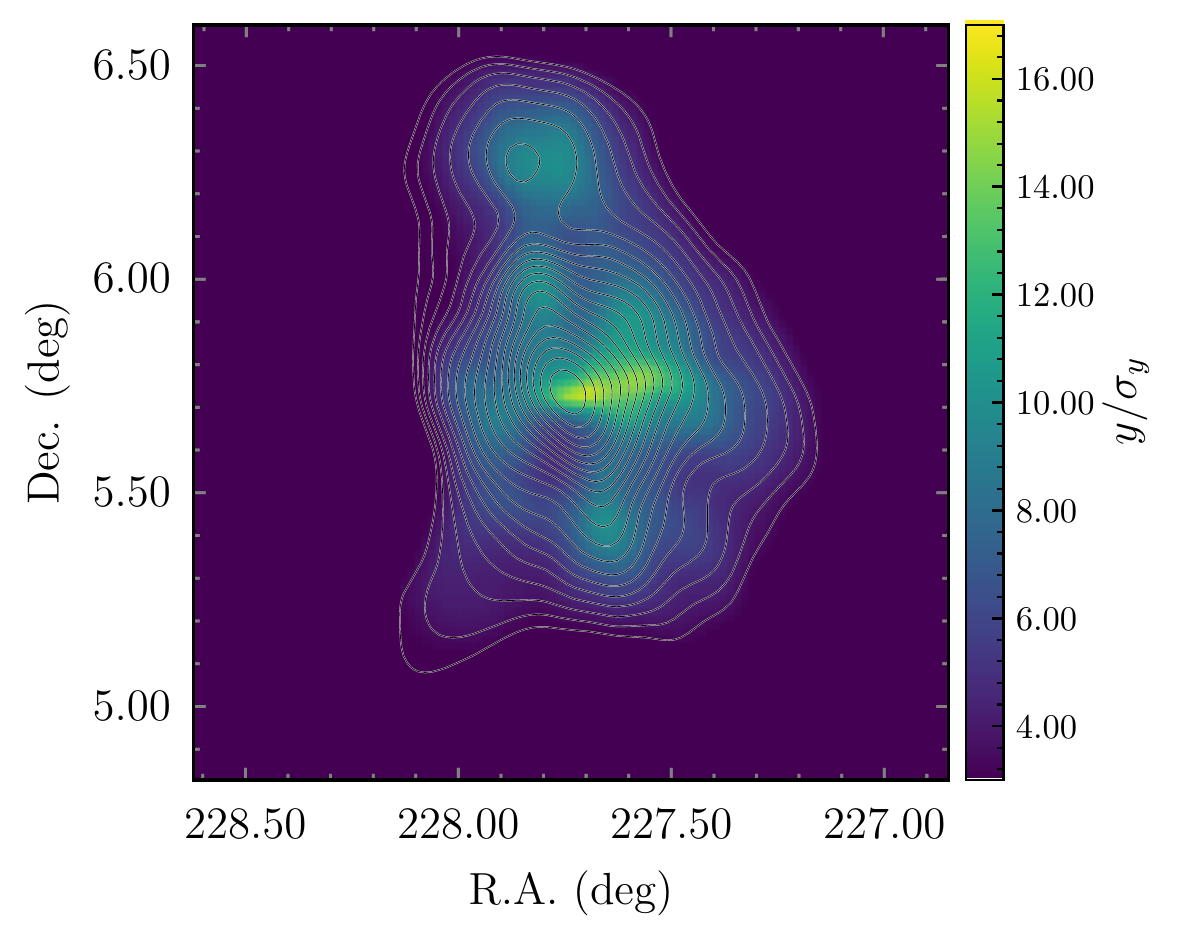}
	\label{fig:a2029_snr}} \quad
	\subfloat[RXCJ1825]{\includegraphics[height=0.23\tht]{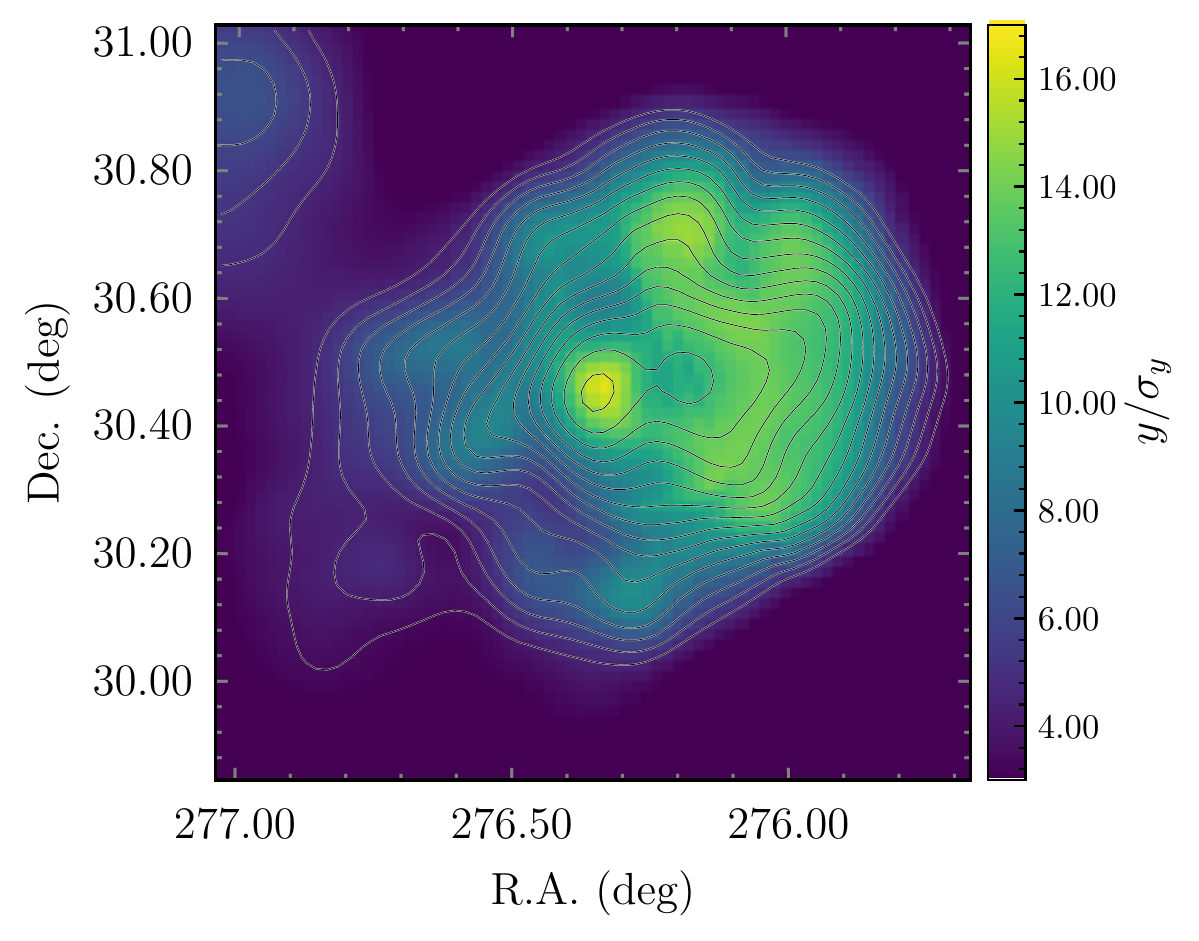}
	\label{fig:rxc1825_snr}}
	\caption{Maps of $ y/\sigma_y $ from a bootstrap run on clusters A2029 and RXCJ1825. The overlapped contours
		represent the tSZ signal.}
\end{figure}

We show in Fig.~\ref{fig:a2029_XSZ} the contours from our map of the tSZ effect, superimposed to the X-ray
surface brightness of \object{A2029}.
We report the data from the \textsl{ROSAT} satellite, which allows the detection of the X-ray emission at larger radii
than those probed by \textsl{Chandra} or \textsl{XMM-Newton}, thanks to its low particle background
\citep{rosat:presentation,vikhlinin:rosatoutskirts}. The raw data have been denoised via the adaptive 
smoothing technique~\citep{eckert:a2029}.
It can be seen that the tSZ signal in the central region matches fairly well with the X-ray surface brightness.
Thanks to the deconvolution, we are able to detect the tSZ emission corresponding to the neighbouring cluster
\object{A2033} with a significance of $ 8 \sigma_y $, which together with A2029 belongs to a small
supercluster \citep{einasto:superclusters}.
The X-ray and the tSZ peaks of A2033 show an offset which is, in any case, smaller than the best resolution of 5 arcmin
provided by \textsl{Planck}.
Our tSZ map clearly highlights an elongated projected structure that connects the two clusters.
As testified by the image showing $ y/\sigma_y $ in Fig.~\ref{fig:a2029_snr}, this elongated excess of signal is
significant to better than 5$ \sigma_y $.
X-ray images do also show such an elongated morphology in the peripheral regions to the north-east, pointing to A2033.
This suggested a possible ongoing merger between the two objects \citep[see e.g.][]{eckert:a2029,walker:a2029}.
However, recent analyses of the reconstructed density field through gravitational lensing,
indicate that this signal is likely due to the gas in the overlapping outskirts of the two clusters at $ R_{200} $,
rather than to a filament connecting them \citep{gonzalez:a2029interacting}.
Our tSZ imaging does represent an improvement with respect to the map shown in fig.~1 of \citet{planck:interactingclusters},
where the emission from A2033 is barely detected, as well as the signal between the two clusters.
However, we cannot favour any of the two hypothesised processes.
On the other hand, our maps may help in discriminating among possible scenarios proposed to explain the observed excess,
for instance in future works combining microwave and X-ray data to model the three-dimensional thermodynamic properties of
the ICM in each component of the system.
\subsubsection{RXCJ1825}
In Fig.~\ref{fig:rxc1825_XSZ}, we show the X-ray surface brightness of \object{RXCJ1825} from \textsl{XMM-Newton},
featuring wavelet denoising. The tSZ contours from our algorithm are superimposed to the image.
As in the case of A2029, the tSZ emission follows pretty well the X-ray surface brightness. In particular,
we notice that the peaks of the signal in the two bands coincide within 1 arcmin.
The tSZ signal shows a significant elongation towards the neighbouring cluster \object{CIZA J1824.1+3029}
(CIZAJ1824 hereafter), even if the peak corresponding to this object is not clearly visible.
A recent analysis based on the kinematics of their member galaxies, suggests that RXCJ1825 and CIZAJ1824 are in a
pre-merger state \citep{girardi:rxc1825}.
Thus, also in this case, our maps may shed a new light on the hypothesis of a possible interaction in future analyses.
Interestingly, our tSZ map shows a $ 14 \sigma_y $-significant elongation to the south-west of the cluster,
highlighted with a green arrow in Fig.~\ref{fig:rxc1825_XSZ}, which agrees with the excess in the X-ray surface brightness
detected with \textsl{XMM-Newton}.
Such an emission may be due to gas stripped from the past interaction between RXCJ1825 and a small, disrupted group of 
galaxies, which have been detected at the same redshift in the optical band \citep{clavico:rxc1825}.
This scenario is also supported by the recent finding of a radio halo extending in the same direction of the X-ray
elongation \citep{botteon:rxc1825radio}.
Another feature we notice in the signal is the presence of a third structure located within $ R_{500} $,
that contributes at a level of $ y \approx \num{5e-5} $, and which is not detected in X-rays.
Nevertheless, the significance of this detection is 35 per cent lower with respect to the signal in the centre,
as demonstrated by the map in Fig.~\ref{fig:rxc1825_snr}, constructed from a bootstrap run. Therefore, its detection is
likely due to some localised irregularity in the instrumental noise in the raw HFI data.

\begin{figure}
	\centering
	\subfloat[A2029]{
	\includegraphics[width=0.35\tw]{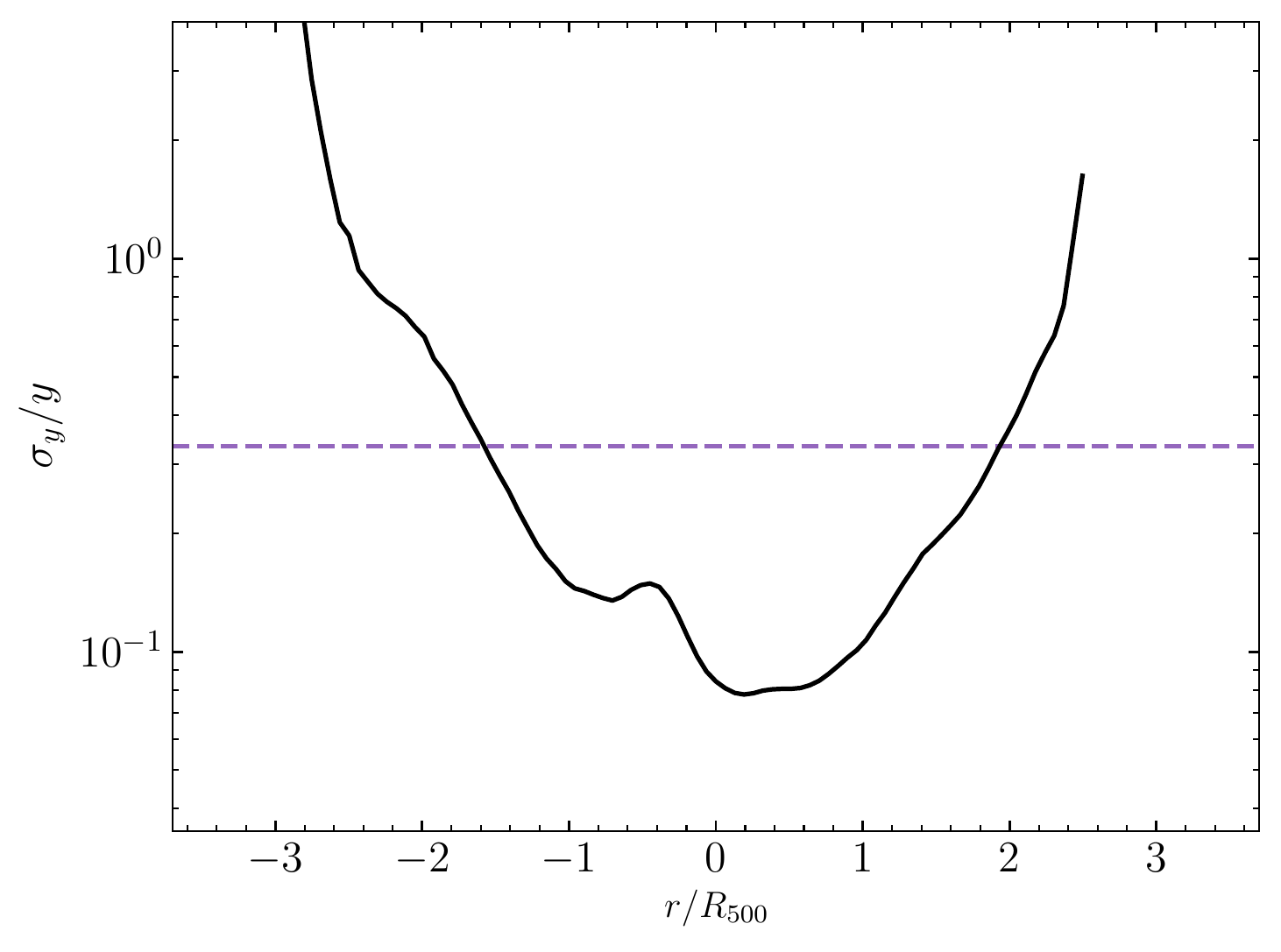}\label{fig:a2029snrcurve}}\qquad
	\subfloat[RXCJ1825]{
	\includegraphics[width=0.35\tw]{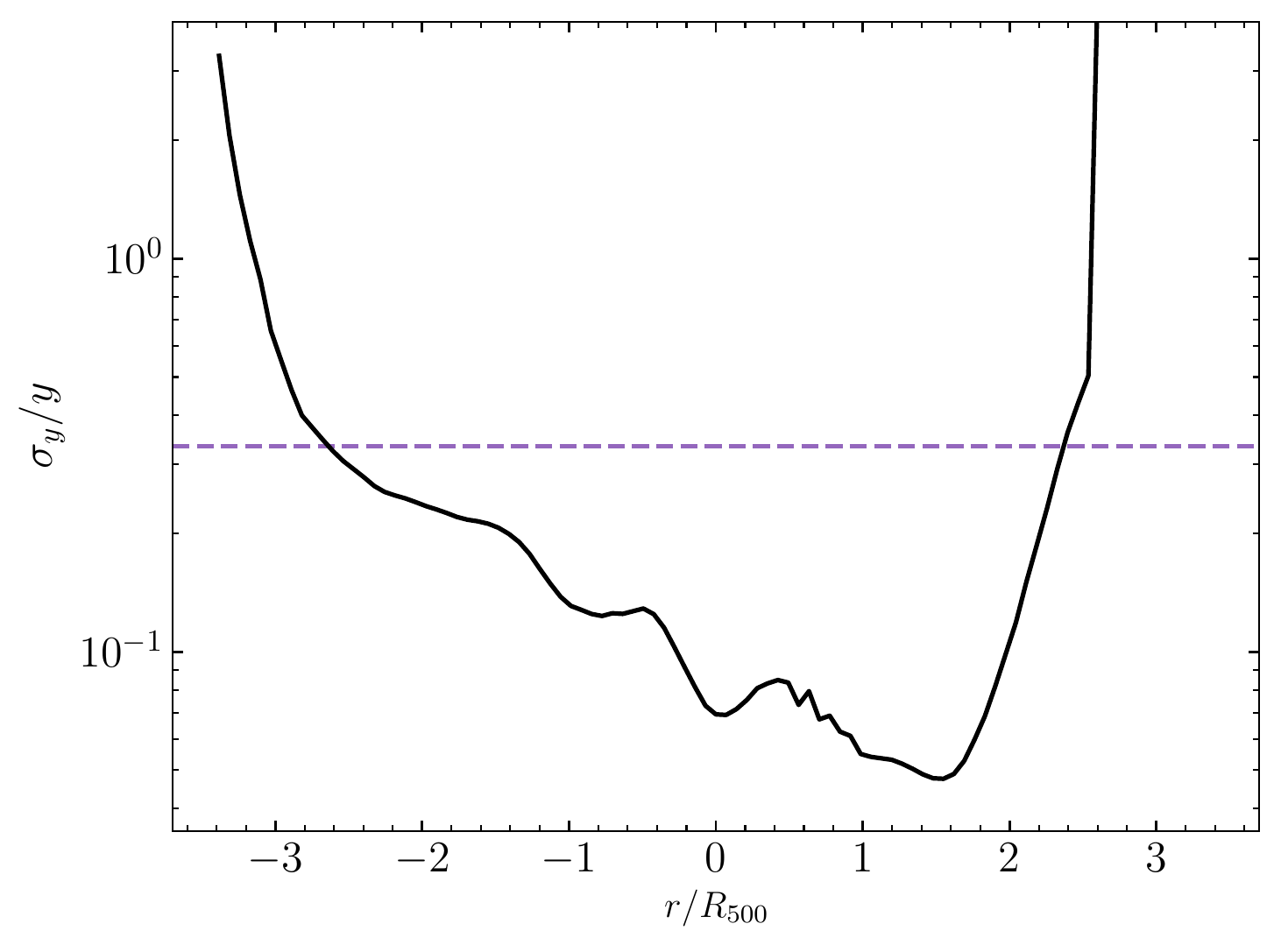}\label{fig:rxc1825snrcurve}}
	\caption{Profiles of $ \sigma_y/y $ of clusters A2029 and RXCJ1825.
		The purple dashed lines correspond to an effective S/N of 3 ($ \sigma_y/y = 1/3 $, see text for details).}
	\label{fig:snr}
\end{figure}

In order to further quantify the significance of the signal reconstructed with our algorithm as a function of the
radius, we computed the profiles of the ratio $ \sigma_y/y $ that is the inverse of the effective S/N.
Specifically, we show in Fig.~\ref{fig:snr} the average of the vertical cuts passing through the centre of the
maps of $ \sigma_y/y $ within a 10 arcmin-wide band, for the clusters A2029 and RXCJ1825.
It can be seen that our algorithm provides a reconstruction of the signal with an effective S/N $ > 3 $
(marked with a dashed purple line in the plots), up to radii $ r \simeq 2R_{500}$ for both clusters.

\section{Conclusions}
\label{sec:conclusions}
We presented maps of the Comptonization parameter of the twelve nearby, massive galaxy clusters
constituting the X-COP sample \citep{xcop:presentation}. We used an improved version of the spectral imaging
algorithm for parametric component separation proposed in \citetalias{bourdin:SZimaging}, featuring wavelet and
curvelet decomposition.
In particular, the enhancements we introduced are: (\emph{i}) a double grey body spectral energy density to model
the emission from thermal dust;
(\emph{ii}) the cleaning of residual contamination from dust and compact point sources;
(\emph{iii}) a new wavelet coefficient-wise deconvolution via beam-weighting in the calculation of the
wavelet transform.
We also illustrated a new method to estimate the error on the reconstructed tSZ signal based on bootstrap extractions
of \textsl{Planck} noise maps.
With this method, we showed it is possible to detect the signal from diffuse filaments and small substructures located
beyond $ R_{500} $. Indeed, we detected interesting features in the outskirts of the majority of the objects in the
sample under study.
In particular, we highlighted diffuse bridges in significant excess with respect to the
background level, which connect the brightest objects of the cluster systems A2029 and RXCJ1825.
These findings are consistent with ancillary surface brightness maps in the X-ray band.
When compared to its original implementation, our new deconvolution technique provided more stability,
an enhanced localisation of the central peak and a more precise reconstruction of the signal in the outskirts.
In particular, the test case of cluster A2319 showed a reduction of contamination from outliers
at radii $ r > 2 R_{500} $, and about 50 per cent lower values of the bootstrap error in the same region.

We plan to use the tSZ maps presented in this work to isolate and mask the signal from substructures located in
cluster outskirts, which will be the topic of a forthcoming paper.
Indeed, this signal may bias-high the reconstructed radial profiles of thermal pressure in the virial region, with
non-negligible consequences on the parameters of the profile, and on estimates of cluster masses relying on the
assumption of hydrostatic equilibrium.\\
Cluster A2319, which is the one detected by \textsl{Planck} with the highest S/N, will be the subject of a dedicated
upcoming work, in which we will highlight in detail the interesting features characterising the tSZ map that
we were able to reconstruct by employing sparse representations.

\begin{acknowledgements}
The authors thank the anonymous referee for useful comments on the manuscript.
They acknowledge the \textsl{Planck} Collaboration for the extraordinary legacy they delivered to the scientific
community. They also acknowledge the use of the image analysis routines of the Interactive Sparse astronomical
data Analysis Package (ISAP) developed in the CosmoStat laboratory at CEA Saclay.
This work makes also an extensive use of \texttt{python}, particularly of the \texttt{astropy} and
\texttt{matplotlib} libraries.
The authors thank Sabrina De Grandi for helpful suggestions.
A.S.B. is grateful to Rocco D'Agostino for useful discussions, and she acknowledges funding from
Sapienza Universit\`{a} di Roma - Progetti per Avvio alla Ricerca Anno 2018, prot. AR11816430D9FBEA.
A.S.B., H.B. and P.M. acknowledge financial contribution from the agreement ASI-INAF n.2017-14-H.0.,
from ASI Grant 2016-24-H.0., and from “Tor Vergata” Grant “Mission: Sustainability” EnClOS (E81I18000130005).
S.E. acknowledges financial contribution from the contracts ASI 2015-046-R.0 and ASI-INAF n.2017-14-H.0.
M.G. is supported by the Lyman Spitzer Jr. Fellowship
(Princeton University) and by NASA Chandra grants GO7-18121X and GO8-19104X.
\end{acknowledgements}
\bibliographystyle{aa}
\bibliography{./BIBLIOGRAPHY.bib}
\end{document}